\documentclass[preprint,showpacs,showkeys,preprintnumbers,amsmath,amssymb,aps]{revtex4}
\usepackage{epsfig}
\usepackage{graphicx}% Include figure files
\usepackage{dcolumn}% Align table columns on decimal point
\usepackage{bm}% bold math
\usepackage{tabularx}
\def\bsub{\begin{subequations}}
\def\esub{\end{subequations}}
\def\be{\begin{eqnarray}}
\def\ee{\end{eqnarray}}

\begin{document}
\baselineskip 0.7cm

\title{Neutrino, Photon Interaction in Unparticle   Physics}
\author{Sukanta Dutta$^1$}\email{Sukanta.Dutta@fnal.gov}
\author{Ashok Goyal$^2$} \email{agoyal@iucaa.ernet.in}
\affiliation{$^1$ Department of Physics and Electronics, SGTB Khalsa College, University of Delhi. Delhi-110007. India.}
\author{  }
\affiliation{$^2$Department of Physics and Astrophysics, University of Delhi. Delhi-110007. India.}

\begin{abstract}
We investigate the impact of unparticle physics on the annihilation of  relic neutrinos with the neutrinos identified as primary source of ultra high energy (UHE) cosmic ray events, producing a cascade of photons and charged particles. We compute the contribution of the unparticle 
exchange to the cross-sections $\nu\,\bar\nu\to\gamma\,\gamma$ and 
$\nu\,\bar\nu\to f\,\bar f$ scattering. We estimate the neutrino photon decoupling temperature from the reaction rate of $\nu\,\bar\nu\to\gamma\,\gamma$. We find  that inclusion of 
unparticles can in fact account for the flux of UHE cosmic rays and can also result in the lowering of neutrino - photon decoupling temperature below the QCD phase transition for unparticle physics parameters in a certain range. We calculate the mean free path of these high energy neutrinos   annihilating themselves with the relic neutrinos to produce  vector and tensor unparticles. 
\end{abstract}

\maketitle

\section{Introduction}
\label{introduction}
High energy neutrino interactions are of great interest in astrophysics, cosmology and in high energy cosmic ray physics. Neutrinos have been considered as possible candidates of Ultra  High energy (UHE) cosmic rays as opposed to protons and photons on account of their ability to travel cosmic galactic distances without significantly degrading their energy. In comparison to protons and photons, neutrinos have relatively weaker interaction cross - section with relic neutrinos and CMBR. By the same token they present difficulty in initiating air showers. Weiler \cite{weiler} proposed that if their energy could correspond to $Z-$  resonance {\it i.e.},
\be
E_\nu\approx \frac{m_Z^2}{2\, m_\nu} \simeq 10^{23} \,\, {\rm eV}\, 
\ee
 they would have significant annihilation cross -section with relic neutrinos of 
 mass consistent with the oscillation data. The difficulty in realizing this scenario is to identify the source of UHE neutrinos with their energy close to $Z-$ resonance. Decay of super heavy relic particles $M_\chi\geq 10^{13}$ GeV has been proposed \cite{gelmini} to be the source of these highly energetic neutrinos that can explain cosmic ray events above the GZK cut off. There however, remains the problem of confining the production of these neutrinos in a spherical shell at red shift 
\be
Z= \left[\frac{M_\chi}{\rm 2\,Eres} -1\right] = \left[\frac{M_\chi\,\, m_\nu}{\rm m^2_Z} -1\right] 
\ee
so that this energy near earth is close to $Z-$ resonance energy  for the $\nu\bar\nu$ annihilation cross section on relic neutrinos to be large. If such a scenario is not realized in nature, we would require large neutrino hadron cross section in the milli barn (mb) region for the neutrinos to initiate showers high in the atmosphere. Current estimate of UHE neutrino - hadron cross section in the standard Model (SM)  by Gandhi et. al. \cite{gandhi} put the cross section in the 10$^{-4}- 10^{-5}$  mb range for $E_\nu \sim 10 ^{21}$ eV. 
\par We thus require anomalously large high energy neutrino interactions. In this context theories of $n-$ extra dimensions with large compactification radius and TeV scale gravity \cite{add} provide the possibility of enhancing neutrino interactions through the exchange of a tower of massive spin 2 bulk gravitons (Kaluza Klien excitations). Contribution of Kaluza Klien excitations and its impact on UHE cosmic ray physics has been discussed by several authors \cite{nussinov} in the literature.

\par The high energy neutrino photon interactions are also of great interest in astrophysics and cosmology. Scattered photons on neutrinos through $\gamma\nu\rightarrow \gamma\nu$ are predominantly circularly polarized due to the left handed nature of neutrinos \cite{dicusrepko}. In the early universe, the photons and neutrinos decouple i.e. the process $\nu\bar\nu\rightarrow \gamma\gamma $ ceases to occur at a temperature $T\sim  1.6$ GeV about one micro second after the big bang. If due to some enhanced neutrino photon interaction,  this temperature can be brought down to QCD phase transition temperature,  some remnants of circular polarization  could perhaps be retained in the cosmic microwave background. In the SM these cross sections are very small, due to the vector and axial vector nature of weak interaction and the leading term for massless neutrinos in fact vanishes due to Yang's theorem \cite{yang}. The cross section for $\gamma\gamma\rightarrow\nu\bar\nu$ has been calculated  in the SM  \cite{abbasabadi} and shows a  $s^3$ behavior upto $W^\pm$  pair production threshold beyond which it starts falling. The cross section is given by
\be
\sigma \bigl(\nu\bar\nu\rightarrow\gamma\gamma \bigr) =\frac{s^3}{20\,\pi} \,\,\left(\frac{A\,\, g^2\,\,\alpha_{\rm em}}{32\,\,\pi\,\,m_W^4} \right)^2 \hskip 0. cm {\rm where}\,\,\, A=14.4\,\,\,.
\ee 

\par The contribution of Kaluza Klien excitations to these processes has been computed by Dicus et. al. \cite{dicuskovner}, who have shown that the contribution is not large enough to allow high energy neutrinos to scatter from relic neutrinos through $\nu\bar\nu\rightarrow \gamma\gamma$ but photon neutrino decoupling temperature may in fact be lowered.
\par Recently Georgi \cite{georgi, georgi1} has proposed that scale invariance, which has been a powerful tool in physics, may indeed exist at a scale much above the TeV scale. He argued that a scale invariant sector with non trivial infrared  fixed point which couples to SM may appear. At low energy scale this gives rise to what has been called an unparticle operator ${\cal O_U}$ with a non integral scale dimension $d_{\cal U}$ having a mass spectrum which looks like a $d_{\cal U}$ number of massless particles. The unparticles have continuous mass spectrum. The unparticle operators can have different Lorentz structures and couple to the SM fields below a large mass scale through an effective non- re-normalizable Lagrangian
\begin{eqnarray}
{\cal L}_{\rm eff.}&=&\frac{{\cal O}_{SM}\,\, {\cal O}_{\cal U}}{M^{d_{\cal U} +d_{SM}-4}}\label{lagran}
\end{eqnarray}
where $\Lambda_{\cal U}$ is energy scale of the order of 1 TeV. It is related to high mass scale $M_{\cal U}$ through 
\be
\kappa ={\cal C_U}\,\,\, \left(\frac{\Lambda_{\cal U}}{M_{\cal U}}\right)^{d_{\cal BZ}+d_{SM}-4}
\ee
where ${\cal C_U} $ is the dimensionless coupling constant, $d_{SM}$ is dimension of SM operator, and  $d_{\cal BZ}$ is the dimension of Banks-Zaks (${\cal BZ}$) scale invariant \cite{bankzak} sector  interacting with SM fields through the exchange of high mass particles $M_{\cal U}$ induced by the Lagrangian given in equation (\ref{lagran}). Unparticle operators with different Lorentz structure corresponding  to scalar, vector, tensor and spinor operators have been considered in the literature. 
\par This unparticle sector can arise as stated in \cite{georgi}  from the hidden sector or from strongly interacting magnetic phase of a specific class of supersymmetric theories \cite{fox} or from hidden valleys model \cite{strassler}. However, we also note that under a specfic conformal invariance \cite{intriligator} the  propagators for vector and tensor are modified.
\par In this paper we study the attenuation of high energy neutrinos through interaction with the present density of relic neutrinos through $\nu\bar \nu\rightarrow {\cal U}$, $\nu\, \bar\nu\rightarrow\gamma\, \gamma $ and $\nu\,\bar\nu\rightarrow f\, \bar f $ processes. The last two processes will proceed through  the exchange of ${\cal U}$ unparticles and would directly produce a cascade of high energy photons and hadrons which could account for the flux of UHE cosmic rays. We also estimate the photon neutrino decoupling temperature. In section \ref{nenebann} we give the unparticle interactions with the SM fields and calculate the cross sections for the above processes. In section \ref{uhesec2} we give an estimate of neutrino - photon decoupling temperature and mean free path of neutrinos through intergalactic journey. This is followed by the discussion of our results in section \ref{discuss}.

\section{Neutrino Antineutrino annihilation}
\label{nenebann}
The effective interactions consistent with SM gauge symmetry for the vector and tensor unparticles with SM fields are given by
\begin{eqnarray}
 \frac{\kappa_1^{\cal V}}{\Lambda_{\cal U}^{d_{\cal U}-1}}\,\, \bar f\,\gamma_\mu \, f\,\,\,  {\cal O_U}^\mu\, ;\,\,\,{\rm and}\,\,\,
 \frac{\kappa_1^{\cal A}}{\Lambda_{\cal U}^{d_{\cal U}-1}}\,\, \bar f\,\gamma_\mu\,\gamma_5 \, f\, {\cal O_U}^\mu\label{veccoup}
\end{eqnarray}
and for tensor unparticles the interactions are
\begin{eqnarray}
\frac{-\,i}{4}\frac{\kappa_T} {\Lambda_{\cal U}^{d_{\cal U}}} \bar f \,\, \bigl( \gamma_\mu \stackrel{\leftrightarrow} {D}_\nu + \gamma_\nu \stackrel{\leftrightarrow} {D}_\mu \bigr) \,\,\psi_f\,\, {\cal O}_{\cal U}^{\mu\nu} \,\, {\rm and}\,\,
\frac{\kappa_T} {\Lambda_{\cal U}^{d_{\cal U}}}
{\cal F}_{\mu\alpha}{\cal F}_{\nu}^\alpha\, {\cal O_U}^{\mu\nu}\label{tenscoup}
\end{eqnarray}
where $f$ stands for a SM fermion doublet or  singlet, ${\cal F}_{\mu\nu}$ is electromagnetic field tensor and the dimensionless coupling constants $\kappa_i$'s are related to the coupling constant ${\cal C_U}$ and the mass scale $M_{\cal U}$ through
\begin{eqnarray}
\frac{\kappa_1^{\cal V,\, A}}{\Lambda_{\cal U}^{d_{\cal U}-1}}=
{\cal C_U^{V,\,A}}\,\, \frac{\Lambda_{\cal U}^{3-d_{\cal U}}}{M^2_{\cal U}}\,\,&{\rm and}&\,\,
\frac{\kappa_0^{{\cal T}}}{\Lambda_{\cal U}^{d_{\cal U}}}=
{\cal C_U}^{{\cal  T}}\,\, \frac{\Lambda_{\cal U}^{2-d_{\cal U}}}{M^2_{\cal U}}\,\,\, .
\end{eqnarray}
The neutrinos being left handed the scalar operator does not couple to them and therefore we consider only the vector and tensor operators. The unparticle propagator for the vector and tensor fields are given by \cite{georgi,cheung1unp}
\be
\bigl[{\cal A_F}(P^2)\bigr]_{\mu\nu}&=&\frac{{\cal A}_{d_{\cal U}}}{2\,\sin\bigl(d_{\cal U}\,\,\pi\bigr)}\,\, \bigl(-\,\, P^2\bigr)^{d_{\cal U}-2}\,\,\pi_{\mu\,\nu}\,(P)\,\,\,\,\, {\rm where}\,\,\, \pi_{\mu\,\nu}\,(P)= -\,g^{\mu\nu}+\frac{P^\mu\,\, P^\nu}{P^2} ; \\
\bigl[{\cal A_F}(P^2)\bigr]_{\mu\nu ,\,\rho\sigma }&=&\frac{{\cal A}_{d_{\cal U}}}{2\,\sin\bigl(d_{\cal U}\,\,\pi\bigr)}\,\, \bigl(-\,\, P^2\bigr)^{d_{\cal U}-2}\,\,T_{\mu\,\nu ,\,\rho\sigma}\,(P)\,\,\,\,\, {\rm where}\nonumber\\
T_{\mu\,\nu ,\,\rho\sigma}\, (P)&=&\frac{1}{2}\left[\pi_{\mu\,\rho}\,(P)\,\, \pi_{\nu\,\sigma}\,(P) +\pi_{\mu\,\sigma}\,(P)\,\, \pi_{\nu\,\rho}\,(P)-\frac{2}{3}\,\, \pi_{\mu\,\nu}\,(P)\,\, \pi_{\rho\,\sigma}\,(P)\right].
\ee
\noindent They satisfy the conditions $P_\mu\,\,\pi^{\mu\nu}\, (P)=0$ and $P_\mu\,\,T^{\mu\nu ,\, \rho\sigma }\, (P)=0$. Further the unparticle operator ${\cal O_U}$ and ${\cal O_U}^{\mu\nu}$ are taken to be Hermitian and transverse and the tensor unparticle operator is also traceless. 
${\cal A}_{d_{\cal U}}$ is the normalization factor for the two point unparticle operator and is given by
\begin{equation}
{\cal A}_{d_{\cal U}}= \frac{16\,\pi^{5/2}}{\left(2\,\pi\right)^{2\,d_{\cal U}}}\,\,\,\frac{\Gamma \left(d_{\cal U}+ \frac{1}{ 2}\right)}
{\Gamma \left(d_{\cal U} -1 \right)\,\,\,\Gamma \left(2\,\,d_{\cal U}\right)}
\end{equation}
The spin averaged cross section induced by the  vector and tensor unparticle operators for the process 
\be
\nu \bigl(p_1\bigr) + \bar\nu \bigl(p_2\bigr) \rightarrow {\cal U} 
\ee
are calculated to be
\be
\sigma_{\rm av.}^{\cal V}\bigl(\nu\bar\nu\rightarrow {\cal U}\bigr)&=& \bigl\vert\kappa_1^{\cal V}\bigr\vert^2 \left(\frac{s}{\Lambda_{\cal U}^2}\right)^{d_{\cal U}-1}\,\, {\cal A}_{d_{\cal U}}\,\,\, \frac{1}{s}\\
\sigma_{\rm av.}^{\cal T}\bigl(\nu\bar\nu\rightarrow {\cal U}\bigr)&=&\frac{1}{32\,\,\Lambda^2_{\cal U}}\,\, 
\bigl\vert\kappa^{\cal T}\bigr\vert^2 \left(\frac{s}{\Lambda_{\cal U}^2}\right)^{d_{\cal U}-1}\,\, {\cal A}_{d_{\cal U}}
\ee
\par For the scattering process 
\be
\nu \bigl(p_1\bigr) + \bar\nu \bigl(p_2\bigr) \rightarrow \gamma_1 \bigl(k_1\bigr) + \gamma_2 \bigl(k_2\bigr) 
\ee
we only have the contribution from tensor unparticle operator. The SM contribution to the above process has been given in reference \cite{dicuskovner}, we can easily compute the total contribution to the spin averaged cross-section and we get
\begin{figure}[bt]
\begin{center}
\vskip -5 cm
\includegraphics[width=20cm,height=30cm]{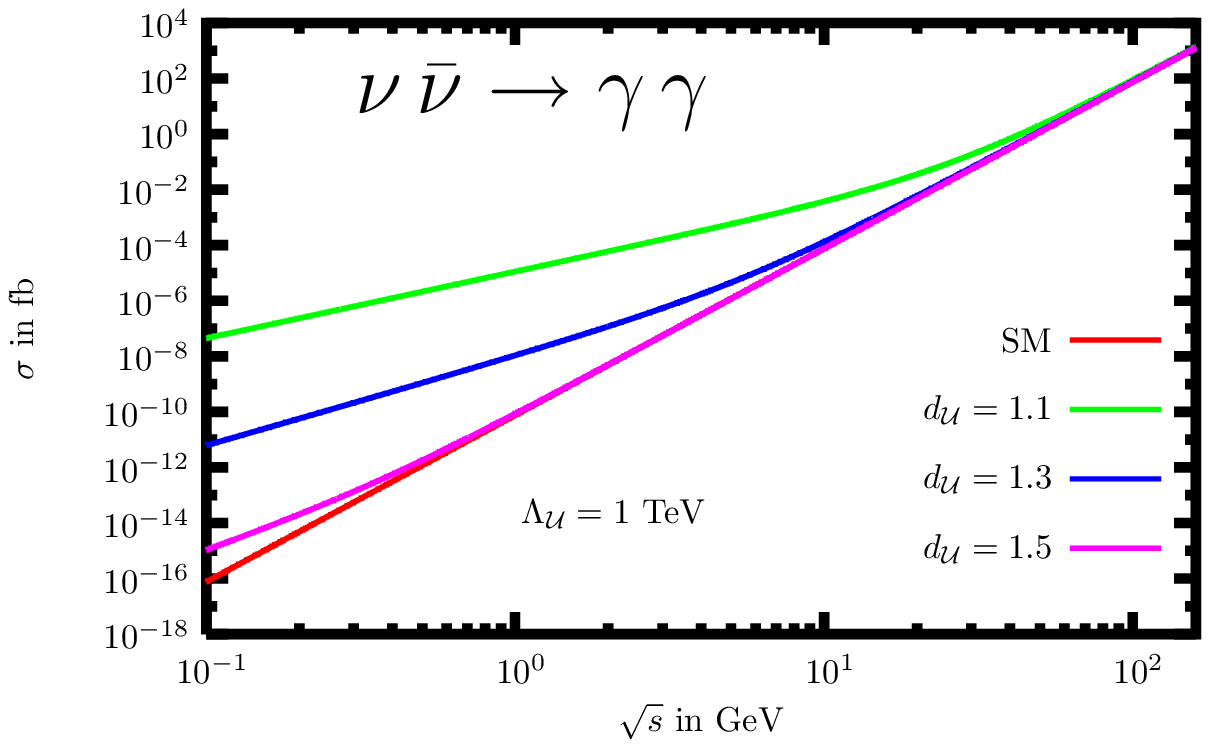}
\vskip -17 cm 
\caption{SM + Unparticle contribution to the total cross section in fb for the photon pair production  from neutrino pair annihilation via tensor unparticle for  $\sqrt{s}$ varying  from 0.1 GeV - 160 GeV }
\label{nunugg}
\end{center}
\end{figure}       
\begin{figure}[bt]
\begin{center}
\vskip -5 cm
\includegraphics[width=20cm,height=30cm]{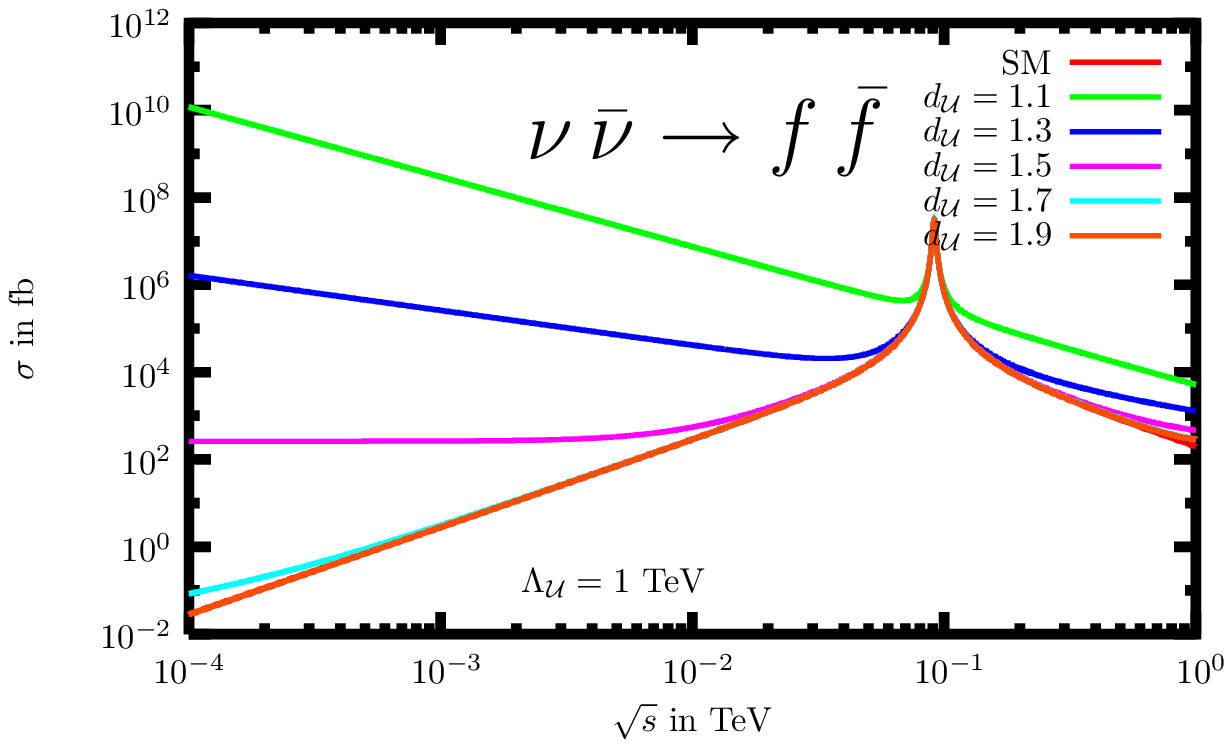}
\vskip -17 cm 
\caption{SM + vector unparticle + tensor unparticle contribution to the total Cross Section in fb for the charged fermion pair production  from neutrino pair annihilation  for $\sqrt{s}$  varying from 0.1 GeV - 1 TeV }
\label{nunuff}
\end{center}
\end{figure}       
\be
\sigma_{\rm av.}\bigl(\nu\bar\nu\rightarrow \gamma\,\gamma\bigr)&=&\frac{1}{20\,\pi}\,\Bigl[{\cal A}_{\rm SM}^2 \, s^3 + 2\, {\cal A}_{\rm SM}\,{\cal A}_{\rm Unp.} \,\,\cos\bigl(\pi\,(d_{\cal U}-2)\bigr)\,\, s^{d_{\cal U}+1}+{\cal A}_{\rm Unp.}^2 \, s^{2\,d_{\cal U}-1}\Bigr] 
\ee
\noindent 
\be
{\rm where}\,\,{\cal A}_{\rm SM}= \left[\frac{14.4\,\,g_W^2\,\,\alpha_{em} }{32\,\,\pi\,\,m_W^4}\right];&& {\cal A}_{\rm Unp.}=\left[\frac{\bigl\vert\kappa^{\cal T}\bigr\vert^2\,\,{\cal Z}_{d_{\cal U}}}{4\,\,\Lambda_{\cal U}^{2\,d_{\cal U}}}\right]\,\, {\rm and} \,\,\, {\cal Z}_{d_{\cal U}}=\frac{
{\cal A}_{d_{\cal U}}} {2\,\sin\bigl(d_{\cal U}\,\,\pi\bigr)}
\ee
\par The cross section for 
\be
\nu \bigl(p_1\bigr) + \bar\nu \bigl(p_2\bigr) \rightarrow f\bigl(p_1^\prime\bigr) + \bar f \bigl(p_2^\prime\bigr) 
\ee
\noindent get contribution from  both the vector and tensor unparticles operators and we get

\be
\sigma_{\rm av.}\bigl(\nu\bar\nu\rightarrow f\,\bar f \bigr) &=&
\frac{2\,G^2_F\,\, s}{3\,\pi}\,\,\bigl\vert{\cal R}(s)\bigr\vert^2\,\,\Biggl[ \bigl({C_{\cal V}^f}^2 +{C_{\cal A}^f}^2\bigr) + 2\,\cos\bigl(\pi\,d_{\cal U}+\Phi\bigr)\, C_{\cal V}^f\,{\cal B}^{\cal V}_{\cal U}+ \left({\cal B}^{\cal V}_{\cal U}\right)^2+ \left({\cal B}^{\cal T}_{\cal U}\right)^2\Biggr]\nonumber\\
{\rm where}&& {\cal B}^{\cal V}_{\cal U}=\left[\frac{\bigl\vert\kappa^{\cal V}\bigr\vert^2\,\, {\cal Z}_{d_{\cal U}}}{2\,\sqrt{2}\,\bigl\vert{\cal R}(s)\bigr\vert }\right]\,\left(\frac{\Lambda^{-2}}{G_{F}}\right)\,\, \left(\frac{s}{\Lambda^2}\right)^{d_{\cal U}-2};\,\,\,
\Phi=\tan^{-1}\left[-\,\,\frac{m_Z\,\Gamma_Z}{s-m_z^2}\right];\nonumber\\
&&{\cal R}(s)=m_z^2\,\frac{(s-m^2_Z)-i\, m_Z\,\Gamma_Z}{(s-m_Z^2)^2+ m_z^2\, \Gamma_Z^2};\,\,\, 
{\cal B}^{\cal T}_{\cal U}=\left[\frac{\sqrt{3}\bigl\vert\kappa^{\cal T}\bigr\vert^2\,\, {\cal Z}_{d_{\cal U}}}{16\,\sqrt{5}\,\bigl\vert{\cal R}(s)\bigr\vert }\right]\,\left(\frac{\Lambda^{-2}}{G_F}\right)\,\, \left(\frac{s}{\Lambda^2}\right)^{d_{\cal U}-1}\nonumber\\
&& C_{\cal V}^f=T_3^f-2\, {\cal Q}_f\,\sin^2\theta_W; \,\,\, {\rm and}\,\,\, C_{\cal A}^f=T^f_3\,\, .
\label{nunuffxsec}
\ee
\noindent In equation (\ref{nunuffxsec}) the second term within the square bracket is the interference term involving  the SM vector and axial current with the vector unparticle current. The interference term  of the  vector and the tensor contribution vanishes identically after angular integration.
\par For a $\nu$ of energy $E_\nu$ annihilating a relic neutrino of mass $m_\nu$ we get
\be
\sigma_{\rm av.}^{\cal V}\bigl(\nu\bar\nu\rightarrow {\cal U}\bigr)&\simeq& 
4\,\,\bigl\vert\kappa_1^{\cal V}\bigr\vert^2 \,\, {\cal A}_{d_{\cal U}}\,\,  10^{8-3\,d_{\cal U}}\,\,\left(2\,\,\frac{m_\nu}{1\, {\rm eV}}\,\, \frac{E_\nu}{10^{21}\, {\rm eV}} \right)^{d_{\cal U}-2} \,\,\left(\frac{\Lambda_{\cal U}}{1\, \, {\rm TeV}}\right)^{2-2\, d_{\cal U}}\,\, {\rm pb}\, ; \\
\sigma_{\rm av.}^{\cal T}\bigl(\nu\bar\nu\rightarrow {\cal U}\bigr)&\simeq& 
12\,\,\bigl\vert\kappa_2^{\cal T}\bigr\vert^2 \,\, {\cal A}_{d_{\cal U}}\,\, 10^{2-3\,d_{\cal U}}\,\,\left( 2\,\, \frac{m_\nu}{1\, {\rm eV}}\,\, \frac{E_\nu}{10^{21}\, {\rm eV}} \right)^{d_{\cal U}-1} \,\,\left(\frac{\Lambda_{\cal U}}{1\, \, {\rm TeV}}\right)^{-\,2\, d_{\cal U}}\,\, {\rm pb}\, ;
\ee
%\noindent The cross sections are shown in the Fig. \ref{nunubV} and \ref{nunubT}.
\be
\sigma_{\rm av.}\bigl(\nu\bar\nu\rightarrow \gamma\, \gamma \bigr)&\simeq&
7.57\times 10^{-5}\,\,\left(2\,\, \frac{m_\nu}{1\, {\rm eV}}\, \frac{E_\nu}{10^{21}\, {\rm eV}} \right)^3 \,\,\,\times\nonumber\\
&& \Biggl[1+ 4.5\times 10^6\,
\bigl\vert\kappa_2^{\cal T}\bigr\vert^2 \,{{\cal Z}_{d _{\cal U}}}\, \,\,\left(2\,\, \frac{m_\nu}{1\, {\rm eV}}\, \frac{E_\nu}{10^{21}\, {\rm eV}} \right)^{ d_{\cal U}-2} \,\left(\frac{\Lambda_{\cal U}}{1\,  {\rm TeV}}\right)^{-\,2\, d_{\cal U}}\nonumber\\
&&+5.09\times 10^{6\,(1-d_{\cal U})}\,\bigl\vert\kappa_2^{\cal T}\bigr\vert^4 \,{{\cal Z}_{d _{\cal U}}}^2\, \,\,\left(2\,\, \frac{m_\nu}{1\, {\rm eV}}\, \frac{E_\nu}{10^{21}\, {\rm eV}} \right)^{2\, d_{\cal U}-4} \,\left(\frac{\Lambda_{\cal U}}{1\,  {\rm TeV}}\right)^{-4\, d_{\cal U}}\Biggr]\, {\rm pb}\, .\nonumber\\
\label{ggxsec}
\ee
%\noindent We plot this cross-section in Figure \ref{nunugg} for varying $\sqrt{s}$ from 100 MeV to 160 GeV (2 $m_W$). We observe that with increasing $d_{\cal U}$, say around $d_{\cal U}$ = 1.7 it converges identically to the SM contribution. 
\be
\sigma_{\rm av.}\bigl(\nu\bar\nu\rightarrow f\,\bar f \bigr)&\simeq& 
1.124\times 10^4\,\, \left(2\,\,\frac{m_\nu}{1 \, {\rm eV}}\,\frac{E_\nu}{10^{21}\,{\rm eV}}\right)\Biggl[ \bigl({C_{\cal V}^f}^2 +{C_{\cal A}^f}^2\bigr)\,\, \left\vert {\cal D}(m_\nu,\, E_\nu)\right\vert^{2}\nonumber\\
&&+\,\, 6.06\, \times\, 10^{4-3\, d_{\cal U}}\,C_{\cal V}^f\,\bigl\vert\kappa^{\cal V}\bigr\vert^2\, {\cal Z}_{d_{\cal U}}\,
\left(2\,\,\frac{m_\nu}{1 \, {\rm eV}}\,\frac{E_\nu}{10^{21}\,{\rm eV}}\right)^{d_{\cal U}-2}\,\,
\left(\frac{\Lambda}{1\, {\rm TeV}}\right)^{2- 2\,d_{\cal U}}\,\nonumber\\
&&\times \left\vert {\cal D}(m_\nu,\, E_\nu)\right\vert\,\, \cos\left\{d_{\cal U}\,\pi + \Theta\right\} \nonumber\\
&&+\,\,9.18\,\times\, 10^{8-\,6\, d_{\cal U}}\,\bigl\vert\kappa^{\cal V}\bigr\vert^4\, {\cal Z}_{d_{\cal U}}^2\,
\left(2\,\,\frac{m_\nu}{1 \, {\rm eV}}\,\frac{E_\nu}{10^{21}\,{\rm eV}}\right)^{2\,d_{\cal U}-4}\,\,
\left(\frac{\Lambda}{1\, {\rm TeV}}\right)^{4- 4\,d_{\cal U}}\nonumber\\
&&+\,\, 1.72\,\times \, 10^{-9-6\,d_{\cal U}}\, \bigl\vert\kappa^{\cal T}\bigr\vert^4\, {\cal Z}_{d_{\cal U}}^2\,
\left(2\,\,\frac{m_\nu}{1 \, {\rm eV}}\,\frac{E_\nu}{10^{21}\,{\rm eV}}\right)^{2\,d_{\cal U}-2}\,\,
\left(\frac{\Lambda}{1\, {\rm TeV}}\right)^{- 2\,d_{\cal U}}\Biggr]\nonumber \\
{\rm  where}&&\nonumber\\
{\cal D}(m_\nu,\, E_\nu)&=&\left[1.2\times 10^{-4}\,\left(2\,\,\frac{m_\nu}{1 \, {\rm eV}}\,\frac{E_\nu}{10^{21}\,{\rm eV}}\right) + 0.027\, i\,\right]^{-1}\,\, {\rm and}\,\,\Theta = {\rm Arg.}\bigl[{\cal D}(m_\nu,\, E_\nu)\bigr].\nonumber\\
\label{ffxec}
\ee
\noindent The cross-section  given in equations  (\ref{ggxsec}) and (\ref{ffxec}) for $\Lambda_{\cal U}=1$ TeV are depicted in Figures \ref{nunugg} and  \ref{nunuff} respectively. 

\section{$\nu$ mean free path and $\nu\,\gamma$ decoupling temperature}
\label{uhesec2}
From the expression of the cross sections given above, we find that the mean free path of the neutrinos in their intergalactic journey is dominated by the annihilation of UHE neutrinos on relic neutrinos ( relic density $n_\nu\simeq 56\,\,{\rm cm}^{-3}$) through the production of unparticles. In the absence of any leptonic asymmetry, we have 
\be
\lambda^V_{\nu\bar\nu\rightarrow {\cal U}}& =&\biggl[ \sigma^{\cal V}\bigl(\nu\,\bar\nu \rightarrow {\cal U}\bigr)\,\,\, n_\nu\biggr]^{-1}\nonumber\\
&\simeq&1.44 \times 10^{3\,d_{\cal U}+1}\,\, \frac{ \bigl\vert \kappa^{\cal V}\bigr\vert^{-2}}{{\cal A}_{d_{\cal U}}}\,\, \left[2\,\,\,\frac{m_\nu}{1\, {\rm eV}}\,\,\frac{E_\nu}{10^{21}\,{\rm eV}}\right]^{2-d_{\cal U}}\,\,\left[\frac{\Lambda_{\cal U}}{1\, {\rm TeV}}\right]^{2\, d_{\cal U}-2}\,\,\, {\rm Mega \,\, Parsec}.\\
&&{\rm and}\nonumber \\
\lambda^T_{\nu\bar\nu\rightarrow {\cal U}}& \simeq&4.81 \times 10^{3\,d_{\cal U}+6}\,\, \frac{ \bigl\vert \kappa^{\cal T}\bigr\vert^{-2}}{{\cal A}_{d_{\cal U}}}\,\, \left[2\,\,\,\frac{m_\nu}{1\, {\rm eV}}\,\,\frac{E_\nu}{10^{21}\,{\rm eV}}\right]^{1-d_{\cal U}}\,\,\left[\frac{\Lambda_{\cal U}}{1\, {\rm TeV}}\right]^{2\, d_{\cal U}}\,\,\, {\rm Mega \,\, Parsec}.
\ee
\noindent The mean free path calculated from  the production of the vector unparticle increases with the with $d_{\cal U}$ and are much less than  the tensor unparticle production as the vector unparticle cross-section dominates over the tensor one. In table \ref{meanfree} we present the mean free paths for various values of $d_{\cal U}$. The corresponding value in the SM, namely $\lambda_{\nu\,\bar\nu\to Z^\star}\simeq 1.3\times 10^{10}$ Mega Parsec. At resonance $\sqrt{s}=m_Z$, the $Z$ exchange gives the largest cross-section with mean free path $\lambda_{\nu\,\bar\nu\to Z^\star}\simeq 0.35\times 10^{5}$ Mega Parsec.
\begin{table}[ht]
\begin{center}

\begin{tabular}{|c|c|c|c|c|c|}\hline\hline
$d_{\cal U}$&1.1 &1.3&1.5&1.7&1.9\\ \hline 
$\lambda^{\cal V}$ in $\bigl\vert\kappa^{\cal V}\bigr\vert^{-2}$ M pc.& 1.28 $\times$ 10$^{5}$ & 3.64 $\times$ 10$^{5}$&2.03 $\times$ 10$^{6}$ & 1.45 $\times$ 10$^{7}$& 1.21 $\times$ 10$^{8}$\\\hline
\end{tabular}
\label{meanfree}
\end{center}
\caption{ Mean free Path of UHE $\nu $ due to annihilation with relic $\nu $ to produce vector  unparticle.} 
\end{table}

 \begin{figure}[bt]
\begin{center}
%\vskip -5 cm
\includegraphics[width=15cm,height=30cm]{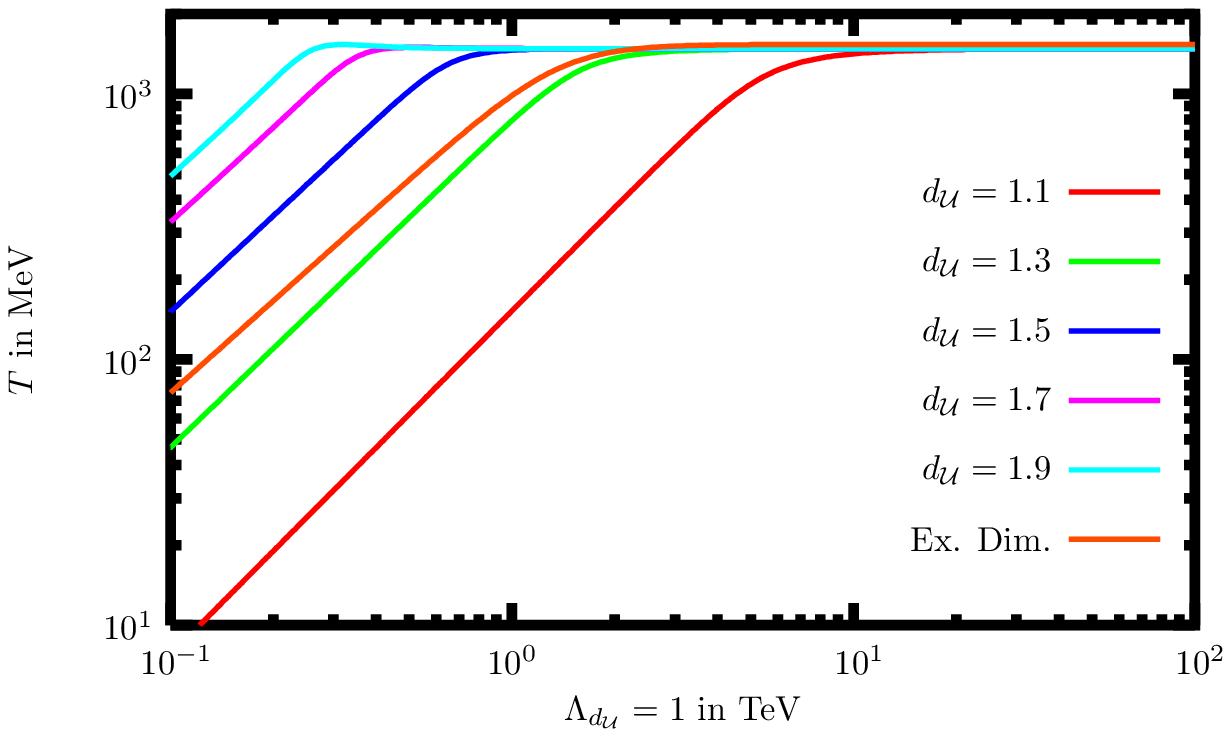}
\vskip -17 cm 
\caption{Contours depicting the decoupling Temperature $T$ in MeV as a function of  $\Lambda_{\cal U}$ for a fixed dimensionless coupling $\bigl\vert \kappa_i\bigr\vert=1$ and various values  of $d_{\cal U}$ varying between 1 and 2. }
\label{decoupfig}
\end{center}
\end{figure}       
\par The temperature at which the reaction $\nu\,\bar\nu\rightarrow \gamma\gamma$ ceases to occur can be obtained from the reaction rate per unit volume i.e.
\be
\Gamma =\frac{1}{\bigl(2\,\pi\bigr)^6} \left[\prod_{i=1}^2 \int\frac{d^3\vec p_i}{e^{E_i/T}+1} \right]\,\, \sigma_{\rm av.}\,\, \bigl\vert \vec v\bigr\vert
\ee
\noindent Substituting the cross sections from equation (\ref{ggxsec})
\be
\Gamma &=& \frac{1}{640\,\pi^5}\Biggl[ \frac{1024}{5}\, {\cal A}_{\rm SM}^2 \, T^{12}\,\, {{\cal F}(5)}^2 +2\,  \, {\cal A}_{\rm SM}\,{\cal A}_{\rm Unp.}\,\,\cos\bigl(\pi\,(d_{\cal U}-2)\bigr)\,\, \frac{4^{d_{\cal U}+3}}{d_{\cal U}+3}\,\,  T^{2\,d_{\cal U}+8}\,\,   {{\cal F}(d_{{\cal U}+3})}^2\nonumber\\
&&\,\,\,\, \,\,\,\,+\frac{4^{2\,d_{\cal U}+1}}{2\,d_{\cal U}+1}\,\, {\cal A}_{\rm Unp.}^2\,\, T^{4\,d_{\cal U}+4}\,\,   {{\cal F}(2\,d_{{\cal U}+1})}^2\Biggr]\nonumber\\
&&{\rm where}\,\,\,\, {\cal F}(n)=\int\frac{x^n}{e^x+1}\,\,dx=\zeta(n+1)\,\Gamma(n+1) \,\,\left(1-\frac{1}{2^n}\right)
\ee

The interaction rate ${\cal R}$ is obtained  by dividing the reaction rate $\Gamma$ by the neutrino number density at temperature $T$ namely 
\be
n_\nu& =&\frac{3\,\zeta\bigl(3\bigr)\, T^3}{4\,\,\pi^2}.
\ee
Thus the reaction rate $R$ is given as
\be
{\cal R}_{\nu\bar\nu\rightarrow\gamma\gamma}&=&1.39\times 10^{-23}\,\, \left(\frac{T}{1\, {\rm MeV}}\right)^9 \,\,\,\otimes \nonumber\\
&&\Biggl[1+ 10^{18-12\,d_{\cal U}}4^{d_{\cal U}} \frac{\cos\bigl(\pi(d_{\cal U}-2)\bigl)}{d_{\cal U}+3}\bigl\vert\kappa^{\cal T}\bigr\vert^2 {\cal Z}_{d_{\cal U}}\,\, {{\cal F}(d_{{\cal U}+3})}^2\,\, \left(\frac{T}{1\, {\rm MeV}}\right)^{2\,d_{\cal U}-4} \,\, \left(\frac{\Lambda_{\cal U} }{1\,{\rm TeV}}\right)^{-\,2\,d_{\cal U}}\nonumber\\
&& + 6.5 \times 10^{38-24\,d_{\cal U}}\,\, 4^{2\, d_{\cal U}}\,\,
\bigl\vert\kappa^{\cal T}\bigr\vert^4\,\, \frac{{\cal Z}^2_{d_{\cal U}}}{(2\,d_{\cal U}+1)}\,\, {{\cal F}(2\,d_{{\cal U}+1})}^2\,\left(\frac{T}{1\,\,  {\rm MeV}}\right)^{4\,d_{\cal U}-8} \left(\frac{\Lambda_{\cal U}}{ 1\, {\rm TeV}}\right)^{-4\, d_{\cal U}}\Biggr]\, {\rm sec}^{-1}\nonumber\\
\ee

\par Using the relation between the age of the universe and the temperature during this era namely 
\be
t_0=2.692\,\, \left[\frac{T}{1\,\, {\rm MeV}}\right]^{-2}\,\, {\rm sec},
\ee
\noindent the condition that at least one interaction takes place gives ${\cal R}_{\nu\bar\nu\rightarrow\gamma\gamma}\,\,\times \, t_0=1$. The solution of this equation gives the decoupling temperature and is shown in Fig. \ref{decoupfig}

\section{Results and Discussion}
\label{discuss}
In order to access the importance of the contribution of the unparticles to ultra high energy neutrino annihilating on cosmic neutrino background, we have
 plotted the average cross-sections $\sigma_{\rm av}\bigl(\nu\,\bar\nu\to\gamma\,\gamma\bigr)$  and $\sigma_{\rm av}\bigl(\nu\,\bar\nu\to f\,\bar f\bigr)$ in Figures \ref{nunugg} and \ref{nunuff}. These cross sections for the highest energy neutrinos have to be large in the vicinity of $\mu$ barns if the high energy photons and charged fermion pairs produced in these reactions which can then fragments into protons, have to account for the flux of the ultra high energy cosmic rays. From these figures we observe that the neutrino annihilation into fermion pairs can indeed be large and may even surpass the cross section at $Z$ resonance which has been considered in the literature \cite{weiler} as a possible mechanism to explain UHE cosmic ray flux for $10^{20}$ eV events. For the highest energy neutrinos $E_\nu\approx 10^{20}\,-10^{21}$ eV and neutrinos of the super Kamiokande motivated mass $\simeq 10^{-2}$ eV, we find the cross-section $\sigma_{\rm av}\bigl(\nu\,\bar\nu\to f\,\bar f\bigr)$ to vary between $10^{9}$ fb to 10  fb for the unparticle dimension $d_{\cal U}$ varying from 1.1 to 1.9 respectively. This cross cross-section is enough for at least one scattering to occur {\it i. e}, they satisfy the condition $\sigma_{\rm av}\bigl(\nu\,\bar\nu\to f\,\bar f\bigr)\,\,\,n_\nu\,\, c\,\, t_0=1$ for $n_\nu\,\, c\,\, t_0\simeq 10^{30}$ cm$^{-2}$ where $n_\nu$ is the relic density (56 cm$^{-3}$) and $t_0=13.5 \times 10^9$ years, the age of the universe. The cross section for the production of a cascade of $\gamma$ rays through $\sigma_{\rm av}\bigl(\nu\,\bar\nu\to\gamma\,\gamma\bigr)$ is not enough to account for the UHE cosmic ray events. However, from Figure \ref{decoupfig} we see that this process can still give enough contribution to significantly lower the neutrino - photon decoupling temperature. 
\par In conclusion, the present study show that unparticle physics can keep alive the hope  of identifying UHE cosmc ray events with the highest energy neutrinos and the possibility of lowering the neutrino - photon decoupling temperature below the QCD phase transition albeit for low unparticle operator dimensions and coupling of the order one for $\Lambda_{\cal U}\approx 1$ TeV.

%%% Local Variables: 
%%% mode: latex
%%% TeX-master: "uhe"
%%% End: 

\end{document}